\documentclass[10pt, a4paper, twocolumn, twoside]{IEEEtran}

\usepackage[utf8]{inputenc}
\usepackage[T1]{fontenc}
\usepackage{url}
\usepackage{ifthen}
\usepackage{cite}
\usepackage[cmex10]{amsmath} % Use the [cmex10] option to ensure complicance
\usepackage{cite}
\usepackage{enumerate}
\usepackage{graphicx}
\usepackage[cmex10]{amsmath} % prevents amsmath from using a Type 3 font for math within footnotes
\usepackage{amssymb}
\usepackage{mathtools}
\usepackage{xspace}
\usepackage{subfigure}
\usepackage[lined,linesnumbered,ruled,commentsnumbered]{algorithm2e} % for algorithm
\usepackage{colonequals}
\usepackage[mathscr]{euscript}
\usepackage{fixltx2e}    % for text subscript
\usepackage{amsthm}
\usepackage{mathrsfs}
\usepackage{hyperref}
\usepackage{caption}

\usepackage{stmaryrd} % for square brackets

\usepackage{xspace}

\usepackage{epsfig}
\usepackage{epstopdf}

\usepackage{tikz}
\usepackage{pgfplots}

\usetikzlibrary{arrows,shapes,backgrounds,plotmarks,positioning}

\pgfplotsset{compat=newest}                         % move axis labels close to the tick label automatically
\pgfplotsset{plot coordinates/math parser=false}
\newlength\figureheight
\newlength\figurewidth

\newtheorem{lemma}{Lemma}

\newtheorem{proposition}{Proposition}

\newtheorem{example}{Example}

\newcommand{\E}{\mathbb{E}}
\newcommand{\Var}{\text{Var}}

\newcommand{\Lat}{\mathcal{L}}

\newcommand{\DP}{\Lat^{\text{DP}}}
\newcommand{\LDP}{\Lat^{\text{LDP}}}
\newcommand{\LogLift}{\Lat^{\text{LL}}}
\newcommand{\SA}{\mathcal{S}}
\newcommand{\X}{\mathcal{X}}
\newcommand{\Y}{\mathcal{Y}}

\newcommand{\DTV}{\text{D}_{\text{TV}}}

\newcommand{\PMat}{\mathbb{P}}
\newcommand{\PV}{\mathbf{P}}
\newcommand{\AMat}{\mathbb{A}}
\newcommand{\AV}{\mathbf{A}}
\newcommand{\Eye}{\mathbb{I}}

\newcommand{\Set}[1]{\{#1\}}

\newcommand{\One}{\mathbf{1}}
\newcommand{\Unit}{\mathbf{e}}

\DeclareMathOperator*{\argmax}{arg\,max}
\DeclareMathOperator*{\argmin}{arg\,min}

\begin{document}
\title{A Linear Reduction Method for Local Differential Privacy and Log-lift}

%%%% Several authors with up to three affiliations:
%\author{%
%  \IEEEauthorblockN{Ni Ding}
%  \IEEEauthorblockA{University of Melbourne\\
%                    ni.ding@unimelb.edu.au}
%  \and
%  \IEEEauthorblockN{Yucheng Liu}
%  \IEEEauthorblockA{The University of Newcastle\\
%                    yucheng.liu@newcastle.edu.au }
%                    \and
%  \IEEEauthorblockN{Farhad Farokhi}
%  \IEEEauthorblockA{University of Melbourne\\
%                    farhad.farokhi@data61.csiro.au }
%}

%\author{Ni~Ding\IEEEauthorrefmark{1}, Yucheng Liu\IEEEauthorrefmark{2}, Farhad Farokhi\IEEEauthorrefmark{1}
%
%\thanks{\IEEEauthorblockA{\IEEEauthorrefmark{1}Ni Ding and Farhad Farokhi are with the University of Melbourne (email: $\{$ni.ding, farhad.farokhi$\}$@unimelb.edu.au).}}
%\thanks{\IEEEauthorblockA{\IEEEauthorrefmark2}Yucheng Liu is with Newcastle University (email: $\{$yucheng.liu$\}$@newcastle.edu.au). }
%}

\author{
    %\IEEEauthorblockN{Ni~Ding\IEEEauthorrefmark{1}, Yucheng~Liu\IEEEauthorrefmark{2}, Farhad~Farokhi\IEEEauthorrefmark{1} and Olya Ohrimenko\IEEEauthorrefmark{1}}
    \IEEEauthorblockN{Ni~Ding\IEEEauthorrefmark{1}, Yucheng~Liu\IEEEauthorrefmark{2} and Farhad~Farokhi\IEEEauthorrefmark{1}}\\
    \IEEEauthorblockA{\IEEEauthorrefmark{1}The University of Melbourne (email: $\{$ni.ding, farhad.farokhi$\}$@unimelb.edu.au).}\\
    \IEEEauthorblockA{\IEEEauthorrefmark{2}The University of Newcastle (email: $\{$yucheng.liu$\}$@newcastle.edu.au). }
}

\maketitle

%%%%%%
%% Abstract:
%% If your paper is eligible for the student paper award, please add
%% the comment "THIS PAPER IS ELIGIBLE FOR THE STUDENT PAPER
%% AWARD." as a first line in the abstract.
%% For the final version of the accepted paper, please do not forget
%% to remove this comment!
%%
\begin{abstract}
This paper considers the problem of publishing data $X$ while protecting the correlated sensitive information $S$. We propose a linear method to generate the sanitized data $Y$ with the same alphabet $\Y = \X$ that attains local differential privacy (LDP) and log-lift at the same time.
It is revealed that both LDP and log-lift are inversely proportional to the statistical distance between conditional probability $P_{Y|S}(x|s)$ and marginal probability $P_{Y}(x)$: the closer the two probabilities are, the more private $Y$ is.
Specifying $P_{Y|S}(x|s)$ that linearly reduces this distance $|P_{Y|S}(x|s) - P_Y(x)| = (1-\alpha)|P_{X|S}(x|s) - P_X(x)|,\forall s,x$ for some $\alpha \in (0,1]$, we study the problem of how to generate $Y$ from the original data $S$ and $X$.
The Markov randomization/sanitization scheme $P_{Y|X}(x|x') = P_{Y|S,X}(x|s,x')$ is obtained by solving linear equations. The optimal non-Markov sanitization, the transition probability $P_{Y|S,X}(x|s,x')$ that depends on $S$, can be determined by maximizing the data utility subject to linear equality constraints on data privacy.
We compute the solution for two linear utility function: the expected distance and total variance distance.
It is shown that the non-Markov randomization significantly improves data utility and the marginal probability $P_X(x)$ remains the same after the linear sanitization method: $P_Y(x) = P_X(x), \forall x \in \X$.
\end{abstract}
\textit{A full version of this paper is accessible at:}
\url{https://arxiv.org/abs/2101.09689}

\section{Introduction}

The privacy-preserving problem can be described as follows. A data curator wants to publish data $X$ that is correlated with the sensitive attribute $S$. To protect privacy, it privatizes $X$ by producing and releasing the sanitized data $Y$. The problem is how to design the sanitization scheme, or the randomization mechanism, to attain a certain level of data privacy.
%
%As to how to measure the privacy, there are two main metics.
We consider two main metrics for measuring privacy: local differential privacy and log-lift.

%-------------------------------------------------------------------------measures---------------------------------------------------------------------------

\emph{Local Differential Privacy}: Let $S$ and $X$ be the random variables in finite alphabets $\SA$ and $\X$, respectively. Assume that the sanitized data has the same alphabet\footnote{We use $\X$ to denote the alphabet of $X$ and $Y$ and $x$, $x'$ or $\tilde{x}$ to denote an instance of either $X$ or $Y$.} as $X$, i.e., $\Y = \X$.
\emph{Differential privacy (DP)} \cite{CalibNoiseDP} measures the statistical distinguishability of $S$. Any sanitization will result in a conditional probability of the output data $P_{Y|S}(x|s)$. The statistical distances of the released data $Y$ conditioned on two adjacent sensitive instances $s, s'$ can be measured by $\DP(S \rightarrow Y) = \max_{x, s,s' \colon s \sim s'} \log \frac{P_{Y|S}(x|s)}{P_{Y|S}(x|s')}$, where $s \sim s'$ denotes $s$ and $s'$ are neighbors that is defined by the Hamming distance constraint $d_{H}(s,s') \leq 1$.
A sanitization mechanism is called $\epsilon$-DP if it generates output $Y$ such that $\DP(S \rightarrow Y) \leq \epsilon$. A small value for $\epsilon$ implies indistinguishability of the sensitive data $S$ when observing the released data $Y$.
%\farhad{[We haven't said anything about the nature/geometry of the set $\SA$ so I don't know how you want to define Hamming distance over it. Are we assuming $\SA$ is binary string, etc? I understand if you were dealing with $\SA^n$ for some $n>1$.]}
%
The local differential privacy (LDP) \cite{LDP2013MiniMax,LDP2014Lalitha} relaxes the neighborhood constraint in DP:
\vspace{-5pt}
\begin{equation} \label{eq:LDP}
	\LDP (S \rightarrow Y) = \max_{x,s,s'} \log \frac{P_{Y|S}(x|s)}{P_{Y|S}(x|s')}. %\vspace{-5pt}
\end{equation}
This is a more general data privacy measure,\footnote{LDP also applies to non-metric space $\SA$, when there is no distance function for the definition of neighborhood, e.g., categorial dataset.} and a stronger notion of privacy: an $\epsilon$-LDP mechanism is always $\epsilon$-DP, but not vice versa.

\emph{Log-lift}: Consider the following statistical inference setting. An adversary wants to infer $S$ from $Y$. The multiplicative difference between the posterior belief $P_{S|Y}(s|x)$ and the prior belief $P_{S}(s)$ denotes the knowledge gain on the sensitive data $S$ by the adversary and therefore indicates the privacy of $Y$.
For a \emph{guessing} adversary, the mutual information $I(S;Y) = \E [\log \frac{P_{S|Y}(s|x)}{P_{S}(s)}]$ and $\log\frac{\E[\max_{s} P_{S|Y}(s|x)]}{\max_{s} P_{S}(s)}$ are used to quantify the average and maximal private information leakage in \cite{PvsInfer2012,Salamatian2015JSTSP} and \cite{Issa2016MaxL,Liu2020ISIT,Liao2017ISITMaxL}, respectively.
They correspond to two extreme cases, $\alpha = 1$ and $\alpha \rightarrow \infty$, of the $\alpha$-leakage proposed in \cite{Liao2019Alpha} based on the Arimoto mutual information $I_{\alpha}^{A}(S;Y)$.
In fact, all these privacy measures can be guaranteed by the \emph{log-lift} \cite{Watchdog2019}:
\begin{equation} \label{eq:LogLift}
    \LogLift( S \rightarrow Y ) = \max_{x,s} \left| \log \frac{P_{S|Y}(s|x)}{P_S(s)} \right|.
\end{equation}
If $\LogLift( S \rightarrow Y) \leq \epsilon$, $I_{\alpha}^{A}(S;Y) \leq \frac{\alpha}{\alpha-1} \epsilon$ for all $\alpha\geq 1$~\cite[Proposition~1]{Watchdog2019}.

While most existing studies only adopt one data privacy measure,\footnote{DP is studied mainly in computer science, where $X = f(S)$ for some deterministic functions $f$ and the privatization usually refers to noise adding mechanism. LDP was originally proposed in \cite{LDP2013MiniMax} for multi-party privacy, where minimax techniques applies to derive fundamental limits on statistical risk assessment and information-theoretic measures. The mutual information, maximal leakage and log-lift are often used in information theory, where $S$ and $X$ are any correlated rvs and the sanitization usually refers to an encoding function.   }
we propose a linear sanitization scheme that attains LDP and log-lift at the same time.
We first reveal that both LDP and log-lift are inverse proportional to the statistical distance between the conditional probability $P_{Y|S}(x|s)$ and the marginal probability $P_{Y}(x)$: the closer these two probabilities are, the more private $Y$ is.
%
%\farhad{[This is very vague. Specially that you have $\alpha$ which is a way of parameterizing $P_{Y|S}(x|s)$, which you don't show here but use it.]}
%
Based on the fact that $P_{Y}(x)$ is the expected value of $P_{X|S}(x|s)$, we request that for all $s,x$ the conditional probability $P_{Y|S}(x|s)$ reduces $P_{X|S}(x|s)$ (in the original dataset) by $\alpha(P_X(x|s) - P_X(x))$ for $\alpha \in (0,1]$.
This ensures a linear decrease $|P_{Y|S}(x|s) - P_Y(x)| = (1-\alpha)|P_{X|S}(x|s) - P_X(x)|,\forall s,x$, which indicates a reduction of approximately a factor of $(1-\alpha)$ in both LDP and log-lift, but remains the same marginal probability: $P_Y(x) = P_X(x), \forall x \in \X$.
%
%
%
%Based on the fact that $P_{Y}(x)$ is the expected value of $P_{Y|S}(x|s)$, we specify  that the conditional probability $P_{Y|S}(x|s)$ of the released data $Y$ linearly reduces the variance $|P_{Y|S}(x|s) - P_Y(x)| = (1-\alpha)|P_{X|S}(x|s) - P_X(x)|,\forall s,x$ for some $\alpha \in (0,1]$. This results in a reduction of approximately a factor of $(1-\alpha)$ in both LDP and log-lift while maintaining the same marginal probability: $P_Y(x) = P_X(x), \forall x \in \X$.
%%
We then determine the randomized scheme that generates such $Y$.
We show that the Markov sanitization scheme $P_{Y|X}(x|x') = P_{Y|S,X}(x|s,x')$ can be obtained by solving linear equations. The optimal non-Markov sanitization scheme, the $P_{Y|S,X}(x|s,x')$ that depends on $S$, can be determined by maximizing the data utility subject to linear equality constraints on data privacy.
We compute the optimal non-Markov sanitization scheme for two linear utility function: the expected distance and total variance distance. The latter is a linear approximation of the mutual information $I(X;Y)$.

\section{Linear Reduction Method for Data Privacy}

We rewrite the maximand in the log-lift \eqref{eq:LogLift} as $|\log \frac{P_{Y|S}(x|s)}{P_Y(x)}|$ and the LDP in \eqref{eq:LDP} as
\begin{equation*}
\begin{aligned}
\LDP (S \rightarrow Y) &= \max_{x,s,s'} \left\{ \log \frac{P_{Y|S}(x|s)}{P_{Y}(x)} + \log \frac{P_{Y}(x)}{P_{Y|S}(x|s')} \right\}\\
                       &= \max_{x,s,s'} \left\{ \log \frac{P_{Y|S}(x|s)}{P_{Y}(x)} - \log \frac{P_{Y|S}(x|s')}{P_{Y}(x)} \right\} .
\end{aligned}
\end{equation*}
%
%
%\[ \LDP (S \rightarrow Y) = \max_{x,s,s'} \left\{ \log \frac{P_{Y|S}(x|s)}{P_{Y}(x)} + \log \frac{P_{Y}(x)}{P_{Y|S}(x|s')} \right\}.  \]
%
Now, both LDP and log-lift are in terms of the conditional probability $P_{Y|S}(x|s)$ and the marginal probability $P_{Y}(x)$, the statistical distance between which is measured by $\log \frac{P_{Y|S}(x|s)}{P_{Y}(x)}$ if $P_{Y|S}(x|s) \geq P_{Y}(x)$ and $-\log \frac{P_{Y|S}(x|s)}{P_{Y}(x)}$ if $P_{Y|S}(x|s) < P_{Y}(x)$. %\footnote{This makes the statistical distance nonnegative, the same as $|\log \frac{P_{Y|S}(x|s)}{P_{Y}(x)}|$.}
In the same way, we can write $\LDP(S \rightarrow X)$ and $\LogLift(S \rightarrow X)$, the LDP and log-lift in the original dataset, in terms of $P_{X|S}(x|s)$ and $P_{X}(x)$.\footnote{For the correlation in the original dataset, denoted by the joint probability $P_{S,X}(s,x)$, we have LDP $\LDP (S \rightarrow X)$ and log-lift $\LogLift( S \rightarrow X )$. They measure the data privacy when $X$ is released without any randomization. This case attains perfect fidelity for the released data with the worst privacy.}

Here, $P_Y(x) = \E[P_{Y|S}(x|\cdot)] = \sum_{s} P_{Y|S}(x|s)P_S(s)$. That is, $P_{Y|S}(x|s)$ can be viewed as a random variable with mean $P_Y(x)$.
Similarly, $P_{X|S}(x|s)$ is a random variable with mean $P_X(x)$.
In this sense, $|\log \frac{P_{X|S}(x|s)}{P_{X}(x)}|$ is a measure of variation. %instance $P_{X|S}(x|s)$.
If it is reduced to a (strictly) smaller variation $|\log \frac{P_{Y|S}(x|s)}{P_{Y}(x)}|$ after the privatized randomization for each $s$, the released data $Y$ is (strictly) more private than the original $X$ in sense of both LDP and log-lift.
To this end, we consider a linear reduction method below.

\subsection{Linear Variance Reduction for Privacy}
\label{sec:Priv}

We set the alphabet of the published data $Y$ the same as $X$: $\Y = \X$. The method generates $Y$ according to the conditional probability:\footnote{It is easy to verify that $0 \leq P_{Y|S}(x|s) \leq 1, \forall s,x$ and $\sum_{x \in \X} P_{Y|S}(x|s) = 1, \forall s$, i.e., $P_{Y|S}(x|s)$ in \eqref{eq:CondProb} is a probability measure. Here, $P_{Y|S}(x|s) =P_{X|S}(x|s)$ if $\alpha = 0$. We consider a strict reduction in LDP and log-lift in this paper and therefore set $\alpha > 0$. }
\begin{align}
		P_{Y|S}(x|s) &= P_{X|S}(x|s) - \alpha (P_{X|S}(x|s) - P_X(x))\nonumber\\
				 &= (1-\alpha) P_{X|S}(x|s) + \alpha P_{X}(x),
				 \label{eq:CondProb}
\end{align}
where $\alpha \in (0,1]$. Here, \eqref{eq:CondProb} is a line search method with $-(P_{X|S}(x|s) - P_X(x))$ being the \emph{descent direction} of the $\ell_1$ distance $|P_{X|S}(x|s) - P_{X}(x)|$ at $P_{X}(x|s)$.
This can also be interpreted as a variation reduction method (see Appendix~\ref{app:VARReduct}).
It is clear that as $\alpha$ increases, $Y$ becomes more private.
For $\alpha=1$, $Y$ is independent of $S$: $P_{Y|S}(x|s) = P_{X}(x)$ for all $s$ and $x$, where \emph{perfect privacy} attains: $\LDP(S \rightarrow Y) = 0$ and $\LogLift(S \rightarrow Y) = 0$.

Eq.~\eqref{eq:CondProb} results in a shift in joint probability $P_{S,Y}(s,x) = P_{Y|S}(x|s) P_{S}(s) = (1-\alpha) P_{S,X}(s,x) + \alpha P_{S}(s) P_{X}(x)$,
%\begin{equation*}
%	\begin{aligned}
%		P_{S,Y}(s,x) &= P_{Y|S}(x|s) P_{S}(s) \\
%				     &= (1-\alpha) P_{S,X}(s,x) + \alpha P_{S}(s) P_{X}(x), \ \ \forall x \in \X,
%	\end{aligned}
%\end{equation*}
but the marginal probability of the released data $Y$ remains the same:
\begin{align}
        P_{Y}(x) & = \sum_{s} P_{S,Y}(s,x) \nonumber\\
                 & = (1-\alpha) P_X(x) + \alpha P_{X}(x) = P_{X} (x), \quad \forall x.
\end{align}
That is, the statistics on the public data $X$ does not change after randomization: the released data $Y$ provides the correct answer to any query on statistical aggregation of $X$.

\subsubsection{Reduction in LDP and Log-lift}

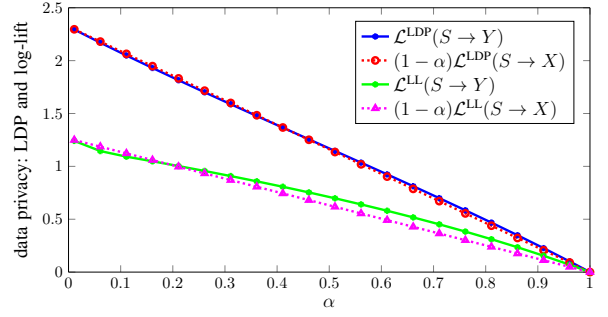
\begin{figure}[tpb]
	\centering
    \scalebox{0.6}{% This file was created by matlab2tikz v0.4.3.
% Copyright (c) 2008--2013, Nico Schlömer <nico.schloemer@gmail.com>
% All rights reserved.
%
% The latest updates can be retrieved from
%   http://www.mathworks.com/matlabcentral/fileexchange/22022-matlab2tikz
% where you can also make suggestions and rate matlab2tikz.
%
%
% defining custom colors
\definecolor{mycolor1}{rgb}{1,0,1}%
\begin{tikzpicture}

\begin{axis}[%
width=4.5in,
height=2.3in,
scale only axis,
separate axis lines,
xmin=0,
xmax=1,
xlabel={\large $\alpha$},
ymin=0,
ymax=2.5,
ylabel={\large data privacy: LDP and log-lift},
legend style={at={(0.55,0.95)},anchor=north west,draw=black,fill=white,legend cell align=left}
]

\addplot [
line width = 1.5pt,
color=blue,
solid,
mark=asterisk,
mark options={solid},
]
table[row sep=crcr]{
0.011 2.29409462637853\\
0.061 2.16986525385814\\
0.111 2.0489393062149\\
0.161 1.93077697955659\\
0.211 1.81490024383404\\
0.261 1.70087909534592\\
0.311 1.58832053550088\\
0.361 1.47685950662016\\
0.411 1.36615122011583\\
0.461 1.2558644476709\\
0.511 1.14567543416287\\
0.561 1.03526214541747\\
0.611 0.924298592275062\\
0.661 0.812448978836068\\
0.711 0.699361407994141\\
0.761 0.584660839446478\\
0.811 0.467940928966434\\
0.861 0.348754273087174\\
0.911 0.22660042413916\\
0.961 0.100910799988354\\
1 0\\
};
\addlegendentry{\large $\LDP(S \rightarrow Y)$ };

\addplot [
line width = 1.5pt,
color=red,
dotted,
mark=o,
mark options={solid},
]
table[row sep=crcr]{
0.011 2.2963868858436\\
0.061 2.18029048109923\\
0.111 2.06419407635487\\
0.161 1.9480976716105\\
0.211 1.83200126686613\\
0.261 1.71590486212176\\
0.311 1.59980845737739\\
0.361 1.48371205263302\\
0.411 1.36761564788866\\
0.461 1.25151924314429\\
0.511 1.13542283839992\\
0.561 1.01932643365555\\
0.611 0.903230028911184\\
0.661 0.787133624166816\\
0.711 0.671037219422448\\
0.761 0.554940814678079\\
0.811 0.438844409933711\\
0.861 0.322748005189343\\
0.911 0.206651600444975\\
0.961 0.090555195700607\\
1 0\\
};
\addlegendentry{\large $(1-\alpha)\LDP(S \rightarrow X)$ };

\addplot [
line width = 1.5pt,
color=green,
solid,
mark=asterisk,
mark options={solid},
]
table[row sep=crcr]{
0.011 1.24098624065362\\
0.061 1.14480759204085\\
0.111 1.09183828929362\\
0.161 1.04810933859242\\
0.211 1.00301339490571\\
0.261 0.956462226619513\\
0.311 0.908358774158753\\
0.361 0.858595931531509\\
0.411 0.807055110119516\\
0.461 0.753604536279995\\
0.511 0.69809722126936\\
0.561 0.640368524762118\\
0.611 0.580233210247547\\
0.661 0.517481859587373\\
0.711 0.451876471728029\\
0.761 0.3831450121334\\
0.811 0.310974597661542\\
0.861 0.235002885236234\\
0.911 0.154807064454491\\
0.961 0.0698896067084003\\
1 0\\
};
\addlegendentry{\large $\LogLift(S \rightarrow Y)$ };

\addplot [
line width = 1.5pt,
color=mycolor1,
dotted,
mark=triangle,
mark options={solid},
]
table[row sep=crcr]{
0.011 1.24914102736962\\
0.061 1.18598930707793\\
0.111 1.12283758678624\\
0.161 1.05968586649455\\
0.211 0.996534146202863\\
0.261 0.933382425911174\\
0.311 0.870230705619484\\
0.361 0.807078985327794\\
0.411 0.743927265036105\\
0.461 0.680775544744415\\
0.511 0.617623824452725\\
0.561 0.554472104161035\\
0.611 0.491320383869346\\
0.661 0.428168663577656\\
0.711 0.365016943285966\\
0.761 0.301865222994277\\
0.811 0.238713502702587\\
0.861 0.175561782410897\\
0.911 0.112410062119208\\
0.961 0.0492583418275179\\
1 0\\
};
\addlegendentry{\large $(1-\alpha)\LogLift(S \rightarrow X)$ };

\end{axis}
\end{tikzpicture}% 
}
	\caption{For the dataset in Example~\ref{ex:main}, the reduction of the LDP $\LDP(S \rightarrow Y)$ and log-lift $\LogLift(S \rightarrow Y)$ as $\alpha$ increases, and their approximations $(1-\alpha) \LDP(S \rightarrow X)$ and $(1-\alpha) \LogLift(S \rightarrow X)$, respectively, in \eqref{eq:PrivAppr}. }
	\label{fig:LinReductionPriv}
\end{figure}

Eq.~\eqref{eq:CondProb} reduces the $\ell_1$-distance by a factor of $1-\alpha$: %\footnote{Similar to the definition of $(0,\delta)$-DP in \cite{Cuff2016DPMI}, we call a mechanism $P_{X|S}$ $(0,\delta)$-LDP if $P_{X|S}(x|s) \leq P_{X|S}(x|s') + \delta, \forall s,s',x$. Eq.~\eqref{eq:CondProb} ensures a strictly smaller $\delta$, that is, $P_{Y|S}(x|s)$ is $(0,(1-\alpha)\delta)$-LDP if $P_{X|S}(x|s)$ is $(0,\delta)$-LDP.This necessarily results in a reduction in the logarithmic statistical distance between $P_{Y|S}(x|s)$ and $P_{Y|S}(x|s')$, e.g., in \eqref{eq:PrivAppr}.}
for each $x$, $|P_{Y|S}(x|s) - P_Y(x)| = (1-\alpha)|P_{X|S}(x|s) - P_X(x)|, \forall s$ and
\begin{multline} \label{eq:L1Norm}
        \quad \Big| \frac{P_{Y|S}(x|s)- P_{{Y|S}}(x|s')}{P_{Y}(x)} \Big| = \\
        (1-\alpha) \Big| \frac{P_{X|S}(x|s) - P_{X|S}(x|s')}{P_{X}(x)} \Big|, \quad \forall s,s'.
\end{multline}
This can be translated to a linear reduction in LDP and log-lift by the first order Taylor approximation $\log(1+x) \approx x$:\footnote{See Appendix~\ref{app:eq:PrivAppr} for the derivation of the approximations in \eqref{eq:PrivAppr}.}
\begin{subequations} \label{eq:PrivAppr}
\begin{align}
    & \LDP(S \rightarrow Y) \approx (1-\alpha) \LDP(S \rightarrow X),  \label{eq:PrivApprLDP} \\
    & \LogLift(S \rightarrow Y) \approx (1-\alpha) \LogLift(S \rightarrow X).
\end{align}
\end{subequations}
%\farhad{[Can you elaborate this more? I don't know how are you getting these.]}
See Fig.~\ref{fig:LinReductionPriv}. The approximations in~\eqref{eq:PrivAppr} are good when $\big| \frac{P_{X|S}(x|s)}{P_X(x)} -1 \big| \leq 1,\forall s,x$.

\section{Optimal Privacy-preserving Scheme}

As explained in Section~\ref{sec:Priv}, one can choose an $\alpha \in (0,1]$ in \eqref{eq:CondProb} to denote a specific privacy level, which results in approximately a reduction of factor $1-\alpha$ in both LDP and Log-lift \eqref{eq:PrivAppr}.
The remaining problem is how to determine a randomized mechanism $P_{Y|X}(x|x')$, which generates $Y$ that holds the private transition probability \eqref{eq:CondProb}. If such mechanism is not unique, we should choose the one that optimizes the data utility.
Denote $U(X;Y)$ the \emph{utility} function that measures the usefulness of the released data $Y$. We consider two types of linear $U(X;Y)$ in this paper: the expected distortion $\E [d(X,Y)]$, where $d(X=x',Y=x) \geq 0 $ and $d(X=x',Y=x) = 0$ for $x = x'$; the total variance distance $\DTV(X,Y) = 1-\sum_{x} P_{X}(x)P_{Y|X}(x|x)$ that measures the expected $\ell_1$ distance between a randomization scheme $P_{Y|X}(x|x')$ and the optimal $P_{Y|X}^*(x|x')$ that maximizes the mutual information $I(X;Y)$. Here, $\DTV(X,Y)$ can be considered as a linear approximation of $I(X;Y)$. See Appendix~\ref{app:DTV}.

The randomized mechanism $P_{Y|X}(x|x')$ can be designed in two ways.
For $S$ being a nesting private attribute of $X$, e.g. $S = f(X)$ for some randomized function $f$ as assumed in \cite{Issa2016MaxL,Liao2019Alpha}, the randomization is conditioned only on the observable data $X$.
In this case, Markov chain $S - X - Y$ forms and the randomized mechanism refers to the Markov transition probability $P_{Y|X}(x|x') = P_{Y|S,X}(x|s,x'), \forall s$, e.g., as in \cite{Ding2019ITW,PF2014}.
If both $S$ and $X$ are observable, e.g., they denote attribute columns in tabular dataset, we can search the optimal randomization over all non-Markov transition probabilities $P_{Y|S,X}(x|s,x')$ \cite[Fig.~1(b)]{Liao2019Alpha}, where $P_{Y|X}(x|x') = P_{Y|S,X}(x|s,x')$ does not necessarily hold for all $s$.

%\farhad{[Should the utility part that follows here go to the beginning of the next subsection? The utility is more general than the Markov case, which is the topic of this subsection.]}

%We show one advantage of adopting the measure $\DTV(X,Y)$ in the next subsection: the problem of maximizing the utility subject to the privacy constraint \eqref{eq:CondProb} can be solved by linear programming (LP).

\subsection{Markov Transition Probability}

The lemma below characterizes the Markov randomization solution.

\begin{lemma} \label{lem:Markov}
The Markov transition probability that satisfies the equality \eqref{eq:CondProb} is
\begin{equation}\label{eq:SolutionMarkov}
	P_{Y|X}(x|x') = \begin{cases} 1-\alpha(1-P_{X}(x)) & x' = x \\
										\alpha P_{X}(x) & x' \neq x \end{cases}. %\IEEEQEDhereeqn
\end{equation}
\end{lemma}
\begin{IEEEproof}
    Lemma holds because
    \begin{multline} \label{eq:SolutionMarkovProof}
        P_{Y|S}(x|s) = \sum_{x'} P_{Y|X}(x|x')P_{X|S}(x'|s) =\\
        (1-\alpha(1- P_X(x))) P_{X|S}(x|s) + \alpha P_{X}(x) \sum_{x'} P_{X|S}(x'|s) = \\
        (1-\alpha)  P_{X|S}(x|s) + \alpha P_X(x), \quad \forall s,x.
    \end{multline}
    The full proof is presented in Appendix~\ref{app:lem:Markov} by solving linear equations.
\end{IEEEproof}
The Markov transition probability in Lemma~\ref{lem:Markov} incurs the expected distortion
$\E [d(X,Y)] = \alpha \sum_{x,x' \colon x' \neq x} P_{X}(x) P_{X}(x') d(X=x',Y=x)$ and the total variance distance $\DTV (X,Y) = \alpha (1-\sum_{x} P_X^2(x))$.

%\begin{equation}
%\begin{aligned}
%        &\quad I(X;Y) \\
%        & = \sum_{x} \sum_{x,x'} P_{Y|X}(x|x') P_{X}(x') \log \frac{P_{Y|X}(x|x')}{P_{Y}(x)} \\
%                 & = \sum_{x} \big( 1-\alpha(1-P_{X}(x)) \big) P_{X}(x) \log \frac{1-\alpha(1-P_{X}(x))}{P_{X}(x)} \\
%                 & \qquad  \alpha \sum_{x} \sum_{x' \colon x' \neq x} P_{X}(x) P_{X}(x') \log\alpha \\
%                 & = \sum_{x} P_{X}(x) \Big[ \big( 1-\alpha(1-P_{X}(x)) \big) \log \frac{1-\alpha(1-P_{X}(x))}{P_{X}(x)} \\
%                 & \qquad  + \alpha (1-P_{X}(x)) \log\alpha \Big]\\
%                 & = H(X) + \\
%                 &\quad \sum_{x} P_{X}(x) \Big( (1-\alpha(1-P_{X}(x)) \log (1-\alpha(1-P_{X}(x)) \\
%                 &\quad + \alpha (1-P_{X}(x)) \log (\alpha P_{X}(x)) ) \Big)\\
%                 &= H(X) + \sum_{x} P_{X}(x) \Big( \log (1-\alpha(1-P_{X}(x)) \\
%                 &\qquad -\alpha(1-P_{X}(x) \log \big( \frac{1-\alpha}{\alpha P_{X}(x)} +1 \big) \Big)
%\end{aligned}
%\end{equation}

\subsection{Non-Markov Transition Probability}

The non-Markov transition probability $P_{Y|S,X}(x|s,x')$ determines a randomized mechanism
\begin{equation} \label{eq:nonMakovPygx}
    P_{Y|X}(x| x') = \sum_{s} P_{Y|S,X}(x|s,x') P_{S|X}(s|x'),
\end{equation}
which is linear in $P_{Y|S,X}(x|s,x')$.
Consider all $P_{Y|S,X}(x|s,x')$ that satisfy \eqref{eq:CondProb}. They are the transition probabilities that attains the same level of privacy (specified by $\alpha$). The problem of searching for an optimal $P_{Y|S,X}^*(x|s,x')$ that maximize the data utility can be formulated as follows. For $\alpha \in (0,1]$,
\begin{subequations} \label{eq:PUT}
\begin{align}
    & \max_{P_{Y|S,X}(x|s,x')} U(X;Y) \label{eq:PUTobj} \\
    &\text{ s.t. } \sum_{x'}P_{Y|S,X}(x|s,x')P_{X|S}(x'|s) = P_{X|S}(x|s)   \nonumber \\
    &\qquad - \alpha (P_{X|S}(x|s) - P_{X}(x)), \quad \forall s,x. \label{eq:PUTconstr}
\end{align}
\end{subequations}
It is clear in \eqref{eq:nonMakovPygx} that the Markov solution is a special case of the non-Makov transition probability. Therefore, the minimizer $P_{Y|S,X}^*(x|s,x)$ of \eqref{eq:PUT} attains a data utility no worse than the Markov solution in Lemma~\ref{lem:Markov} in general.
See Example~\ref{ex:main}.
Since the constraints in \eqref{eq:PUTconstr} are linear, problem \eqref{eq:PUT} is concave maximization if $U(X;Y)$ is concave in $P_{Y|X}(x|x')$.\footnote{The concavity does not hold for general $U(X;Y)$. For example, the mutual information $I(X;Y)$ is convex in $P_{Y|S,X}(x|s,x')$.}
For linear function $U(X;Y)$, \eqref{eq:PUTobj} can be formulated by linear programming (LP).

Below, we compute the solutions for the utility functions $U(X;Y) = -\DTV(X;Y)$ and $U(X;Y) = -\E [d(X,Y)]$. The proofs of Propositions~\ref{prop:DTV} and \ref{prop:ED} are in Appendix~\ref{app:prop:DTV_ED}.

\begin{proposition} \label{prop:DTV}
    For $U(X;Y) = -\DTV(X;Y)$, the solution to problem \eqref{eq:PUT} is any transition probability $P_{Y|S,X}^*(x|s, x')$ satisfying the followings for each $s$:
    \begin{subequations} \label{eq:SolutionNonMarkov}
        \begin{align}
        &P_{Y|S,X}^*  (x| s,x) = \min \left\{  1 - \alpha \Big( 1 - \frac{P_{X}(x)}{P_{X|S}(x|s)} \Big), 1 \right\}, \label{eq:SolutionNonMarkov1} \\
%        &= \begin{cases}
%            1 & x \in \X^+(s) \\
%            1 - \alpha \left( 1 - \frac{P_{X}(x)}{P_{X|S}(x|s)} \right) & x \in \X^-(s)
%            \end{cases},\label{eq:SolutionNonMarkov1}\\
        &P_{Y|S,X}^*(x | s,x') = 0 , \quad \forall  x \in \X^-(s), x' \in \X \colon x'\neq x, \label{eq:SolutionNonMarkov2} \\
        &P_{Y|S,X}^*(x | s,x') = 0 , \quad \forall  x' \in \X^+(s), x \in \X \colon x\neq x',  \label{eq:SolutionNonMarkov3}\\
        &\sum_{x' \in \X^-(s)} P_{Y|S,X}^*(x|s,x') P_{X|S}(x'|s) \nonumber \\
        & \qquad = -\alpha \big( P_{X|S}(x|s) - P_{X}(x) \big), \quad \forall x \in \X^+(s),  \label{eq:SolutionNonMarkov4} \\
        &\sum_{x \in \X^+(s)} P_{Y|S,X}^*(x|s,x')\nonumber \\
        &\qquad\qquad   = \alpha \big( 1-\frac{P_{X}(x')}{P_{X|S}(x'|s)} \big), \quad  x' \in \X^-(s), \label{eq:SolutionNonMarkov5}
        \end{align}
    \end{subequations}
    where $\X^+(s)= \Set{x \in \X \colon P_{X}(x) \geq P_X(x|s)}$ and $\X^-(s) = \Set{x \in \X \colon P_{X}(x) < P_{X|S}(x|s)}$. \hfill \IEEEQED
\end{proposition}

 \begin{figure*}[t]%[htbp]
   \centerline{
        \subfigure{\scalebox{0.6}{% This file was created by matlab2tikz v0.4.3.
% Copyright (c) 2008--2013, Nico Schlömer <nico.schloemer@gmail.com>
% All rights reserved.
%
% The latest updates can be retrieved from
%   http://www.mathworks.com/matlabcentral/fileexchange/22022-matlab2tikz
% where you can also make suggestions and rate matlab2tikz.
%
\begin{tikzpicture}

\begin{axis}[%
width=4.5in,
height=2.3in,
scale only axis,
separate axis lines,
xmin=0,
xmax=4.35,
xlabel={\large privacy: $\LDP(S \rightarrow Y)$},
ymin=0,
ymax=1.4,
ylabel={\large utility loss: $\DTV(X;Y)$},
legend style={at={(0.3,0.95)},anchor=north west,draw=black,fill=white,legend cell align=left}
]
\addplot [
color=blue,
%only marks,
%mark=*,
%mark options={solid},
line width = 1.5pt,
]
table[row sep=crcr]{
4.33369550070895 0.015047725\\
4.24171529947679 0.028727475\\
4.15411029218673 0.042407225\\
4.07042468932958 0.0560869750000001\\
3.99026918652724 0.069766725\\
3.91330860084346 0.083446475\\
3.8392522528695 0.097126225\\
3.76784639381165 0.110805975\\
3.69886817697875 0.124485725\\
3.63212081044728 0.138165475\\
3.5674296235946 0.151845225\\
3.50463884820796 0.165524975\\
3.44360896380666 0.179204725\\
3.38421449247841 0.192884475\\
3.32634215484395 0.206564225\\
3.26988931839604 0.220243975\\
3.21476268426027 0.233923725\\
3.16087716969339 0.247603475\\
3.10815495229085 0.261283225\\
3.05652464858145 0.274962975\\
3.00592060492418 0.288642725\\
2.95628228274221 0.302322475\\
2.90755372339359 0.316002225\\
2.85968308058161 0.329681975\\
2.81262221029724 0.343361725\\
2.76632630997291 0.357041475\\
2.72075359989612 0.370721225\\
2.67586504104918 0.384400975\\
2.63162408445818 0.398080725\\
2.58799644788962 0.411760475\\
2.54494991635917 0.425440225\\
2.50245416343706 0.439119975\\
2.46048059076914 0.452799725\\
2.41900218359697 0.466479475\\
2.37799338036669 0.480159225\\
2.33742995477548 0.493838975\\
2.29728890882374 0.507518725\\
2.25754837562772 0.521198475\\
2.2181875309065 0.534878225\\
2.17918651219338 0.548557975\\
2.14052634493861 0.562237725\\
2.10218887477089 0.575917475\\
2.06415670527186 0.589597225\\
2.02641314069278 0.603276975\\
1.98894213310756 0.616956725\\
1.95172823355298 0.630636475\\
1.914756546756 0.644316225\\
1.87801268909088 0.657995975\\
1.84148274944666 0.671675725\\
1.80515325271794 0.685355475\\
1.76901112566102 0.699035225\\
1.73304366488263 0.712714975\\
1.69723850675046 0.726394725\\
1.66158359903451 0.740074475\\
1.62606717410521 0.753754225\\
1.59067772352948 0.767433975\\
1.55540397391892 0.781113725\\
1.5202348638963 0.794793475\\
1.48515952205643 0.808473225\\
1.45016724580647 0.822152975\\
1.4152474809786 0.835832725\\
1.38038980211438 0.849512475\\
1.34558389332639 0.863192225\\
1.31081952964695 0.876871975\\
1.27608655877843 0.890551725\\
1.24137488316262 0.904231475\\
1.20667444228927 0.917911225\\
1.17197519516605 0.931590975\\
1.13726710287345 0.945270725\\
1.10254011112852 0.958950475\\
1.06778413278197 0.972630225\\
1.03298903017214 0.986309975\\
0.99814459725839 0.999989725\\
0.963240541454642 1.013669475\\
0.928266465081313 1.027349225\\
0.893211846350746 1.041028975\\
0.858066019797314 1.054708725\\
0.822818156058728 1.068388475\\
0.787457240909524 1.082068225\\
0.751972053441229 1.095747975\\
0.716351143276236 1.109427725\\
0.680582806693792 1.123107475\\
0.64465506153662 1.136787225\\
0.608555620755444 1.150466975\\
0.572271864435817 1.164146725\\
0.535790810137041 1.177826475\\
0.499099081356329 1.191506225\\
0.462182873912447 1.205185975\\
0.425027920021575 1.218865725\\
0.387619449813607 1.232545475\\
0.349942150009272 1.246225225\\
0.311980119446592 1.259904975\\
0.27371682110892 1.273584725\\
0.235135030265198 1.287264475\\
0.196216778285513 1.300944225\\
0.15694329164033 1.314623975\\
0.11729492552893 1.328303725\\
0.0772510915100778 1.341983475\\
0.0367901784241615 1.355663225\\
0 1.367975\\
};
\addlegendentry{\large Markov solution in Lemma~\ref{lem:Markov}};

\addplot [
color=red,
%only marks,
%mark=*,
%mark options={solid},
line width = 1.5pt,
]
table[row sep=crcr]{
4.33369550070895 0.00514800000000006\\
4.24171529947679 0.00982800000000008\\
4.15411029218673 0.0145080000000001\\
4.07042468932958 0.0191880000000001\\
3.99026918652724 0.023868\\
3.91330860084346 0.028548\\
3.8392522528695 0.033228\\
3.76784639381165 0.037908\\
3.69886817697875 0.0425880000000001\\
3.63212081044728 0.047268\\
3.5674296235946 0.051948\\
3.50463884820796 0.0566280000000001\\
3.44360896380666 0.061308\\
3.38421449247841 0.0659880000000001\\
3.32634215484395 0.0706680000000001\\
3.26988931839604 0.0753480000000001\\
3.21476268426027 0.0800280000000001\\
3.16087716969339 0.0847080000000001\\
3.10815495229085 0.0893880000000001\\
3.05652464858145 0.094068\\
3.00592060492418 0.0987480000000001\\
2.95628228274221 0.103428\\
2.90755372339359 0.108108\\
2.85968308058161 0.112788\\
2.81262221029724 0.117468\\
2.76632630997291 0.122148\\
2.72075359989612 0.126828\\
2.67586504104918 0.131508\\
2.63162408445818 0.136188\\
2.58799644788962 0.140868\\
2.54494991635917 0.145548\\
2.50245416343706 0.150228\\
2.46048059076914 0.154908\\
2.41900218359697 0.159588\\
2.37799338036669 0.164268\\
2.33742995477548 0.168948\\
2.29728890882374 0.173628\\
2.25754837562772 0.178308\\
2.2181875309065 0.182988\\
2.17918651219338 0.187668\\
2.14052634493861 0.192348\\
2.10218887477089 0.197028\\
2.06415670527186 0.201708\\
2.02641314069278 0.206388\\
1.98894213310756 0.211068\\
1.95172823355298 0.215748\\
1.914756546756 0.220428\\
1.87801268909088 0.225108\\
1.84148274944666 0.229788\\
1.80515325271794 0.234468\\
1.76901112566102 0.239148\\
1.73304366488263 0.243828\\
1.69723850675046 0.248508\\
1.66158359903451 0.253188\\
1.62606717410521 0.257868\\
1.59067772352948 0.262548\\
1.55540397391892 0.267228\\
1.5202348638963 0.271908\\
1.48515952205643 0.276588\\
1.45016724580647 0.281268\\
1.4152474809786 0.285948\\
1.38038980211438 0.290628\\
1.34558389332639 0.295308\\
1.31081952964695 0.299988\\
1.27608655877843 0.304668\\
1.24137488316262 0.309348\\
1.20667444228927 0.314028\\
1.17197519516605 0.318708\\
1.13726710287345 0.323388\\
1.10254011112852 0.328068\\
1.06778413278197 0.332748\\
1.03298903017214 0.337428\\
0.99814459725839 0.342108\\
0.963240541454642 0.346788\\
0.928266465081313 0.351468\\
0.893211846350746 0.356148\\
0.858066019797314 0.360828\\
0.822818156058728 0.365508\\
0.787457240909524 0.370188\\
0.751972053441229 0.374868\\
0.716351143276236 0.379548\\
0.680582806693792 0.384228\\
0.64465506153662 0.388908\\
0.608555620755444 0.393588\\
0.572271864435817 0.398268\\
0.535790810137041 0.402948\\
0.499099081356329 0.407628\\
0.462182873912447 0.412308\\
0.425027920021575 0.416988\\
0.387619449813607 0.421668\\
0.349942150009272 0.426348\\
0.311980119446592 0.431028\\
0.27371682110892 0.435708\\
0.235135030265198 0.440388\\
0.196216778285513 0.445068\\
0.15694329164033 0.449748\\
0.11729492552893 0.454428\\
0.0772510915100778 0.459108\\
0.0367901784241615 0.463788\\
0 0.468\\
};
\addlegendentry{\large non-Markov solution in Proposition~\ref{prop:DTV}};

\end{axis}
\end{tikzpicture}% 
}}     \qquad \qquad
        \subfigure{\scalebox{0.6}{% This file was created by matlab2tikz v0.4.3.
% Copyright (c) 2008--2013, Nico Schlömer <nico.schloemer@gmail.com>
% All rights reserved.
%
% The latest updates can be retrieved from
%   http://www.mathworks.com/matlabcentral/fileexchange/22022-matlab2tikz
% where you can also make suggestions and rate matlab2tikz.
%
\begin{tikzpicture}

\begin{axis}[%
width=4.5in,
height=2.3in,
scale only axis,
separate axis lines,
xmin=0,
xmax=4.35,
xlabel={\large privacy: $\LDP(S \rightarrow Y)$},
ymin=0,
ymax=2,
ylabel={\large utility loss: $H(X) - I(X;Y)$},
legend style={at={(0.4,0.95)},anchor=north west,draw=black,fill=white,legend cell align=left}
]

\addplot [
color=blue,
solid,
line width = 1.5pt,
]
table[row sep=crcr]{
4.33369550070895 0.0750638319364914\\
4.24171529947679 0.129213903722219\\
4.15411029218673 0.178177080253951\\
4.07042468932958 0.223683925109126\\
3.99026918652724 0.266581167273402\\
3.91330860084346 0.307374580926939\\
3.8392522528695 0.346401116200573\\
3.76784639381165 0.383901515871391\\
3.69886817697875 0.420056412289487\\
3.63212081044728 0.45500633952596\\
3.5674296235946 0.488863735149282\\
3.50463884820796 0.521720579901843\\
3.44360896380666 0.553653494025826\\
3.38421449247841 0.584727268231645\\
3.32634215484395 0.614997387895766\\
3.26988931839604 0.644511885650178\\
3.21476268426027 0.673312731907908\\
3.16087716969339 0.701436898979188\\
3.10815495229085 0.728917189267854\\
3.05652464858145 0.755782889501288\\
3.00592060492418 0.782060294389572\\
2.95628228274221 0.807773130731019\\
2.90755372339359 0.832942904536484\\
2.85968308058161 0.857589187866738\\
2.81262221029724 0.881729857911094\\
2.76632630997291 0.905381297834068\\
2.72075359989612 0.928558566722291\\
2.67586504104918 0.951275544337358\\
2.63162408445818 0.973545055159684\\
2.58799644788962 0.99537897528203\\
2.54494991635917 1.01678832500066\\
2.50245416343706 1.03778334940164\\
2.46048059076914 1.05837358880947\\
2.41900218359697 1.07856794062616\\
2.37799338036669 1.09837471381913\\
2.33742995477548 1.11780167710104\\
2.29728890882374 1.13685610167013\\
2.25754837562772 1.15554479923939\\
2.2181875309065 1.17387415596701\\
2.17918651219338 1.19185016280644\\
2.14052634493861 1.20947844271623\\
2.10218887477089 1.22676427510491\\
2.06415670527186 1.2437126178323\\
2.02641314069278 1.26032812704297\\
1.98894213310756 1.27661517506993\\
1.95172823355298 1.29257786661374\\
1.914756546756 1.30822005337515\\
1.87801268909088 1.32354534729581\\
1.84148274944666 1.33855713254144\\
1.80515325271794 1.35325857634466\\
1.76901112566102 1.36765263880947\\
1.73304366488263 1.38174208176652\\
1.69723850675046 1.39552947675688\\
1.66158359903451 1.40901721221186\\
1.62606717410521 1.42220749988773\\
1.59067772352948 1.43510238060626\\
1.55540397391892 1.44770372934485\\
1.5202348638963 1.46001325971369\\
1.48515952205643 1.47203252785157\\
1.45016724580647 1.48376293576647\\
1.4152474809786 1.49520573414213\\
1.38038980211438 1.50636202462699\\
1.34558389332639 1.51723276161729\\
1.31081952964695 1.52781875354171\\
1.27608655877843 1.53812066365033\\
1.24137488316262 1.5481390103066\\
1.20667444228927 1.55787416677589\\
1.17197519516605 1.56732636049973\\
1.13726710287345 1.57649567183979\\
1.10254011112852 1.58538203226966\\
1.06778413278197 1.59398522198712\\
1.03298903017214 1.60230486691217\\
0.99814459725839 1.610340435029\\
0.963240541454642 1.61809123202119\\
0.928266465081313 1.62555639613962\\
0.893211846350746 1.6327348922309\\
0.858066019797314 1.6396255048409\\
0.822818156058728 1.64622683029159\\
0.787457240909524 1.65253726761079\\
0.751972053441229 1.65855500817148\\
0.716351143276236 1.66427802386951\\
0.680582806693792 1.66970405363549\\
0.64465506153662 1.67483058803478\\
0.608555620755444 1.67965485165883\\
0.572271864435817 1.68417378294713\\
0.535790810137041 1.68838401099871\\
0.499099081356329 1.69228182882992\\
0.462182873912447 1.69586316240387\\
0.425027920021575 1.69912353458661\\
0.387619449813607 1.70205802296066\\
0.349942150009272 1.7046612101291\\
0.311980119446592 1.7069271247403\\
0.27371682110892 1.70884917091328\\
0.235135030265198 1.71042004297521\\
0.196216778285513 1.71163162133392\\
0 1.71203342891352\\
};
\addlegendentry{\large Markov solution in Lemma~\ref{lem:Markov}};

\addplot [
color=red,
solid,
line width = 1.5pt,
]
table[row sep=crcr]{
4.33369550070895 0.0240910544986719\\
4.24171529947679 0.0427334734293185\\
4.15411029218673 0.0607921964534357\\
4.07042468932958 0.0791208183004946\\
3.99026918652724 0.0843558506602351\\
3.8392522528695 0.107740196160246\\
3.76784639381165 0.124369789976749\\
3.69886817697875 0.143599871254866\\
3.63212081044728 0.155762150209583\\
3.5674296235946 0.179027224605532\\
3.44360896380666 0.18265546994868\\
3.38421449247841 0.186246281050308\\
3.32634215484395 0.205490339571884\\
3.26988931839604 0.209056581846505\\
3.21476268426027 0.211712369521154\\
3.16087716969339 0.227639807166173\\
3.05652464858145 0.250309007660948\\
2.90755372339359 0.261646703484109\\
2.81262221029724 0.271000020947103\\
2.72075359989612 0.295904541802285\\
2.67586504104918 0.30984285420892\\
2.50245416343706 0.318419990117638\\
2.33742995477548 0.365809046624513\\
2.29728890882374 0.393856534994926\\
2.25754837562772 0.40078799760834\\
2.14052634493861 0.413251962711704\\
1.69723850675046 0.441178918413847\\
1.55540397391892 0.481219300567538\\
1.5202348638963 0.485803312372056\\
1.24137488316262 0.519832501051159\\
1.17197519516605 0.528087610109153\\
0.963240541454642 0.531893236061635\\
0.387619449813607 0.539405468590246\\
0.196216778285513 0.547662364722248\\
0.11729492552893 0.573503009084026\\
0 0.618035718346873\\
};
\addlegendentry{\large non-Markov solution in Proposition~\ref{prop:DTV}};

\end{axis}
\end{tikzpicture}% 
}} }
   \caption{The privacy-utility tradeoff obtained from the dataset in Example~\ref{ex:main} by enumerating $\alpha \in (0,1]$: for each value of $\alpha$, the conditional probability $P_{Y|S}(x|s)$ in \eqref{eq:CondProb} is determined, where we get the privacy measure $\LDP(S \rightarrow Y)$ and
   obtain the Markov randomization schemes in Lemma~\ref{lem:Markov} and Proposition~\ref{prop:DTV}, respectively. We plot the resulting total variance distance $\DTV(X;Y)$ and the utility loss in terms of mutual information $H(X) - I(X;Y)$ vs. $\LDP(S \rightarrow Y)$. The non-Markov solution outperforms Markov solution. }
   \label{fig:PUT_LDPDTVED}
 \end{figure*}

We can directly determine the optimal $P_{Y|S,X}^*(x|s,x')$ by Proposition~\ref{prop:DTV}: for each $s$, do the assignments in \eqref{eq:SolutionNonMarkov1}-\eqref{eq:SolutionNonMarkov3};
determine $P_{Y|S,X}(x|s,x')$ for all $x' \in \X^-(s)$ and $x \in \X^+(s)$ by solving linear equations formed by \eqref{eq:SolutionNonMarkov4} and \eqref{eq:SolutionNonMarkov5}.
Here, \eqref{eq:SolutionNonMarkov1} in fact saturates the diagonal entry $P_{Y|S,X}^*(x| s,x)$ for each $x \in \X$ in the constrained set\footnote{This can be seen from the proof of Proposition~\ref{prop:DTV}: the diagonal entry $P_{Y|S,X}^*  (x| s,x)$ cannot be increased any further without breaching the constraint \eqref{eq:PUTconstr}. }
and the solution to the linear equations \eqref{eq:SolutionNonMarkov4} and \eqref{eq:SolutionNonMarkov5} is not unique.\footnote{This is because the dimension of the null space formed by \eqref{eq:SolutionNonMarkov4} and \eqref{eq:SolutionNonMarkov5} is no less than 1. See the explanation in Appendix~\ref{app:NullSpace}.}

Proposition below shows that when the expected distance is used for utility measure, the problem can be reduced to an LP with reduced dimension of decision variables that is constrained by \eqref{eq:SolutionNonMarkov4} and \eqref{eq:SolutionNonMarkov5}.

%To complete the optimal solution \eqref{eq:SolutionNonMarkov} for each $s \in \SA$, we still need to determine the entries of $P_{Y|S,X}(x|s,x')$ for all $x \in \X^-(s)$ and $x' \in \X^+(s)$. This can be done as follows.
%%
%From \eqref{eq:PUTconstr}, \eqref{eq:SolutionNonMarkov1} and \eqref{eq:SolutionNonMarkov2}, we have
%%
%\begin{multline} \label{eq:SolutionNonMarkov4}
%    \sum_{x' \in \X^-(s)} P_{Y|S,X}^*(x|s,x') P_{X|S}(x'|s) = \\
%    -\alpha \big( P_{X|S}^*(x|s) - P_{X}(x) \big)
%\end{multline}
%for all $x \in \X^+(s)$ and
%\begin{equation} \label{eq:SolutionNonMarkov5}
%    \sum_{x \in \X^+(s)} P_{Y|S,X}(x|s,x') = \alpha \big( 1-\frac{P_{X}(x')}{P_{X|S}(x'|s)} \big)
%\end{equation}
%for all $x' \in \X^-(s)$.
%%

\begin{proposition} \label{prop:ED}
    For $U(X;Y) = -\E [d(X,Y)]$, the solution to problem \eqref{eq:PUT} is the transition probability $P_{Y|X}^*(x|x,s)$ that holds \eqref{eq:SolutionNonMarkov}. The minimizer of
\begin{equation}
\begin{aligned} \label{eq:SolutionNonMarkovED}
    & \min \sum_{x \in \X^+(s),x' \in \X^-(s)} P_{Y|S,X}(x|s,x') P_{S,X}(s,x')) d(x',x)\\
    & \text{s.t. } \eqref{eq:SolutionNonMarkov4} \text{ and } \eqref{eq:SolutionNonMarkov5}.
\end{aligned}
\end{equation}
determines $P_{Y|S,X}^*(x|s,x')$ for all $x \in \X^+(s)$ and $x' \in \X^-(s)$ for each $s$.
\end{proposition}

In problem~\eqref{eq:PUT}, the data privacy constraint \eqref{eq:PUTconstr} is strengthen by increasing $\alpha$, while the maximal utility decreases. Therefore, the privacy utility tradeoff (PUT) can be obtained by varying $\alpha \in (0,1]$.

\begin{figure}[tpb]
	\centering
    \scalebox{0.6}{% This file was created by matlab2tikz v0.4.3.
% Copyright (c) 2008--2013, Nico Schlömer <nico.schloemer@gmail.com>
% All rights reserved.
%
% The latest updates can be retrieved from
%   http://www.mathworks.com/matlabcentral/fileexchange/22022-matlab2tikz
% where you can also make suggestions and rate matlab2tikz.
%
\begin{tikzpicture}

\begin{axis}[%
width=4.5in,
height=2.3in,
scale only axis,
xmin=0,
xmax=2.8,
xlabel={\large privacy: $\LogLift(S \rightarrow Y)$},
ymin=0,
ymax=3,
ylabel={\large utility loss: $\E[d(X,Y)]$},
legend style={at={(0.3,0.95)},anchor=north west,draw=black,fill=white,legend cell align=left}
]
\addplot [
color=blue,
solid,
line width = 1.5pt,
]
table[row sep=crcr]{
2.76652060363487 0.0298755875\\
2.6842375380685 0.0570352125\\
2.60639528747453 0.0841948375\\
2.53253895652013 0.1113544625\\
2.46228015340553 0.1385140875\\
2.3952846266833 0.1656737125\\
2.33126264787008 0.1928333375\\
2.26996143907865 0.2199929625\\
2.2111591450692 0.2471525875\\
2.15465998649821 0.2743122125\\
2.1002903270563 0.3014718375\\
2.04789545520376 0.3286314625\\
1.99733693014179 0.3557910875\\
1.94849037732184 0.3829507125\\
1.90124364510792 0.4101103375\\
1.85549525383909 0.4372699625\\
1.81115308334099 0.4644295875\\
1.76813325620297 0.4915892125\\
1.72635918279409 0.5187488375\\
1.68576074069718 0.5459084625\\
1.64627356647727 0.5730680875\\
1.6078384418212 0.6002277125\\
1.57040075934923 0.6273873375\\
1.53391005600352 0.6545469625\\
1.49831960400754 0.6817065875\\
1.46358605107754 0.7088662125\\
1.42966910293662 0.7360258375\\
1.39653124229976 0.7631854625\\
1.36413747941502 0.7903450875\\
1.33245513000188 0.8175047125\\
1.30145361705348 0.8446643375\\
1.27110429348991 0.8718239625\\
1.24138028308447 0.8989835875\\
1.21225633744905 0.9261432125\\
1.19428467319503 0.9533028375\\
1.18171492931738 0.9804624625\\
1.16903470597432 1.0076220875\\
1.15624204385399 1.0347817125\\
1.14333493105639 1.0619413375\\
1.13031130119437 1.0891009625\\
1.11716903140823 1.1162605875\\
1.10390594028895 1.1434202125\\
1.09051978570513 1.1705798375\\
1.07700826252812 1.1977394625\\
1.06336900024968 1.2248990875\\
1.04959956048591 1.2520587125\\
1.03569743436088 1.2792183375\\
1.02166003976301 1.3063779625\\
1.00748471846668 1.3335375875\\
0.993168733110948 1.3606972125\\
0.978709264026912 1.3878568375\\
0.964103405904465 1.4150164625\\
0.949348164288564 1.4421760875\\
0.934440451894478 1.4693357125\\
0.919377084730671 1.4964953375\\
0.904154778017114 1.5236549625\\
0.888770141885976 1.5508145875\\
0.873219676850582 1.5779742125\\
0.85749976902751 1.6051338375\\
0.841606685095505 1.6322934625\\
0.825536566973608 1.6594530875\\
0.80928542619954 1.6866127125\\
0.792849137987834 1.7137723375\\
0.776223434945579 1.7409319625\\
0.759403900421821 1.7680915875\\
0.742385961464694 1.7952512125\\
0.725164881358177 1.8224108375\\
0.707735751708017 1.8495704625\\
0.69009348404372 1.8767300875\\
0.672232800900677 1.9038897125\\
0.654148226343313 1.9310493375\\
0.635834075886677 1.9582089625\\
0.617284445770078 1.9853685875\\
0.598493201532103 2.0125282125\\
0.579453965831709 2.0396878375\\
0.56016010545489 2.0668474625\\
0.540604717440659 2.0940070875\\
0.520780614253757 2.1211667125\\
0.500680307924364 2.1483263375\\
0.480295993067258 2.1754859625\\
0.459619528684026 2.2026455875\\
0.438642418642177 2.2298052125\\
0.417355790713991 2.2569648375\\
0.395750374045688 2.2841244625\\
0.373816474913712 2.3112840875\\
0.351543950609441 2.3384437125\\
0.328922181276241 2.3656033375\\
0.305940039503102 2.3927629625\\
0.282585857456954 2.4199225875\\
0.258847391310621 2.4470822125\\
0.234711782694926 2.4742418375\\
0.210165516871122 2.5014014625\\
0.185194377282985 2.5285610875\\
0.159783396105932 2.5557207125\\
0.13391680036245 2.5828803375\\
0.107577953118117 2.6100399625\\
0.0807492892092099 2.6371995875\\
0.053412244880101 2.6643592125\\
0.0255471806245223 2.6915188375\\
0 2.7159625\\
};
\addlegendentry{\large Markov solution in Lemma~\ref{lem:Markov}};

\addplot [
color=red,
solid,
line width = 1.5pt,
]
table[row sep=crcr]{
2.76652060363487 0.0171212849298598\\
2.6842375380685 0.0290091395305359\\
2.60639528747453 0.0482508939646591\\
2.53253895652013 0.0536639362912734\\
2.46228015340553 0.0658194847956911\\
2.3952846266833 0.0779247315367046\\
2.33126264787008 0.0899984892028917\\
2.26996143907865 0.102051188682575\\
2.2111591450692 0.114089097256271\\
2.15465998649821 0.1261163576282\\
2.1002903270563 0.138135322561067\\
2.04789545520376 0.150148028382414\\
1.99733693014179 0.162155799485032\\
1.94849037732184 0.174159609113376\\
1.81115308334099 0.2101538479289\\
1.72635918279409 0.226044067524311\\
1.68576074069718 0.237878742886357\\
1.64627356647727 0.258121662513305\\
1.6078384418212 0.261548072949441\\
1.53391005600352 0.294086133926874\\
1.49831960400754 0.306072815205971\\
1.46358605107754 0.308886823383359\\
1.39653124229976 0.332556212423418\\
1.36413747941502 0.354010739990431\\
1.33245513000188 0.365994569179189\\
1.30145361705348 0.377982546705319\\
1.27110429348991 0.389961158124506\\
1.24138028308447 0.391729746993702\\
1.21225633744905 0.403564458642672\\
1.19428467319503 0.425908809284224\\
1.18171492931738 0.437890858793911\\
1.15624204385399 0.46185431717263\\
1.14333493105639 0.47383575591416\\
1.13031130119437 0.474636591868004\\
1.11716903140823 0.497798115348678\\
1.09051978570513 0.521759852815174\\
1.07700826252812 0.533740510142594\\
1.06336900024968 0.533823921201523\\
1.03569743436088 0.557415799959436\\
1.02166003976301 0.581661890241037\\
0.993168733110948 0.592919967540376\\
0.978709264026912 0.604754692114048\\
0.934440451894478 0.640258863266605\\
0.904154778017114 0.66392831158243\\
0.873219676850582 0.687597761054185\\
0.85749976902751 0.699432486662883\\
0.841606685095505 0.711267211936756\\
0.825536566973608 0.73739540741353\\
0.792849137987834 0.746771387599888\\
0.725164881358177 0.794110291073015\\
0.707735751708017 0.805945020574003\\
0.654148226343313 0.841449200043886\\
0.617284445770078 0.865118655722788\\
0.520780614253757 0.941027571377144\\
0.500680307924364 0.953005341412408\\
0.480295993067258 0.964983049843424\\
0.373816474913712 1.00713539193346\\
0.328922181276241 1.03080484728204\\
0.305940039503102 1.04263957314763\\
0.258847391310621 1.06630902929322\\
0.234711782694926 1.09673380078442\\
0.159783396105932 1.11364794042295\\
0.053412244880101 1.16098685268828\\
0 1.27347283405394\\
};
\addlegendentry{\large non-Markov solution in Proposition~\ref{prop:ED}};

\end{axis}
\end{tikzpicture}% 
}
	\caption{The privacy-utility tradeoff in terms of the expected distance $\E[d(X;Y)]$ vs. the log-lift $\LogLift(S \rightarrow Y)$ obtained from the dataset in Example~\ref{ex:main}.}
	\label{fig:PUT_LDPED}
\end{figure}
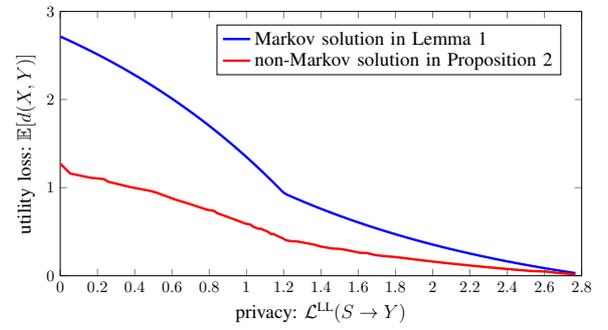

\begin{example} \label{ex:main}
	 Consider an database with the joint probability $P_{X|S}(x|s)$ below.
    \begin{table}[h]
        \begin{center}
        \begin{tabular}{ccccc}
            \hline\hline
            & $X = a$ & $X=b$ & $X=c$ & $X=d$\\ \hline
            $S = 1$ & 0.2  & 0.1 & 0.5 & 0.2 \\ \hline
            $S = 2$   & 0.5  & 0.3 & 0.1 & 0.1 \\  \hline
            \end{tabular}
        \end{center}
    \end{table}
    The marginal probabilities are $P_S(1)=0.3$, $P_S(2)=0.7$, $P_{X}(a) =0.41$, $P_{X}(b) =0.24$, $P_{X}(c) =0.22$ and $P_{X}(d) =0.13$. For $\alpha = 0.5$, we show how to obtain the optimal transition probability $P_{Y|S,X}^*(x|s,x')$ in Proposition~\ref{prop:DTV}.
    For $S = 1$, $\X^+(1)= \Set{a,b}$ and $X^-(1) = \Set{c,d}$. By \eqref{eq:SolutionNonMarkov1}, we set $P_{Y|S,X}^*(a|1,a) = P_{Y|S,X}^*(b|1,b) = 1$, $P_{Y|S,X}^*(c|1,c) = 0.72$ and $P_{Y|S,X}^*(d|1,d) = 0.825$.
    We obtain one solution to the linear equations \eqref{eq:SolutionNonMarkov4} and \eqref{eq:SolutionNonMarkov5}: $P_{Y|S,X}^*(a|1,c) = 0.21$, $P_{Y|S,X}^*(b|1,c) = 0.07$, $P_{Y|S,X}^*(a|1,d) = 0$ and $P_{Y|S,X}^*(b|1,d) = 0.175$.
    All other entries of $P_{Y|S,X}^*(x|1,x')$ are set to $0$. The transition probability $P_{Y|S,X}^*(x|2,x')$ for all $x,x'$ can be determined in the same way. Apply $P_{Y|X}^*(x| x') = P_{Y|S,X}^*(x|1,x') P_{S|X}(1|x') + P_{Y|S,X}^*(x|2,x') P_{S|X}(2|x')$ by \eqref{eq:nonMakovPygx}. The resulting mutual information is $I(X;Y) = 1.19$ and $\DTV(X;Y)= 0.23$.
    They can be compared to the Markov solution in Lemma~\ref{lem:Markov}, where we get $I(X;Y) = 0.36$ and $\DTV(X;Y)= 0.68$.
    This means that when a certain level of data privacy is guaranteed, adopting non-Markov randomization can significantly improve the utility. This can also be seen in Fig.~\ref{fig:PUT_LDPDTVED} and \ref{fig:PUT_LDPED}.

    We then obtain the solution in Proposition~\ref{prop:ED} to the problem \eqref{eq:PUT} for $U(X;Y) = -\E [d(X,Y)]$. The procedure is the same as above, except that $P_{Y|S,X}^*(x|s,x')$ for all $x \in \Set{a,b}$ and $x' \in \Set{c,d}$ is determined by solving the minimization \eqref{eq:SolutionNonMarkovED} for all $s$. The resulting privacy-utility tradeoff is shown in Fig.~\ref{fig:PUT_LDPED}.

\end{example}

\section{Conclusion}

Noting that $P_{X}(x)$ is the expected value of the conditional probability $P_{X|S}(x|s)$ w.r.t. the marginal probability of $S$, we proposed a privacy-preserving method that generates the sanitized data $Y$ with $P_{Y|S}(x|s)$ that linearly reduces the variance of $P_{X|S}(x|s)$ in the original data $X$. This randomization method maintains the marginal probability $P_{Y}(x) = P_{X}(x), \forall x$.
We showed that $\LDP(S \rightarrow Y)$ and $\LogLift(S \rightarrow Y)$ can be expressed in terms of $|\log \frac{P_{Y|S}(x|s)}{P_Y(x)}|$ and therefore the proposed method reduces both LDP and log-lift. Specifically, $\LDP(S \rightarrow Y) \approx (1-\alpha) \LDP(S \rightarrow X)$ and $\LogLift(S \rightarrow Y) \approx (1-\alpha) \LogLift(S \rightarrow X)$, where $\alpha \in (0,1]$ can be considered as the privacy level. We considered Markov and non-Markov sanitization schemes to generate $Y$. While the Markov scheme was obtained by solving linear equations, we formulated an LP to compute the optimal non-Markov scheme for two linear utility functions. The experimental results showed that the non-Markov scheme significantly improves data utility.

There are two aspects that can be further explored.
While the proposed linear method reduces the variance of $P_{X|S}(x|s)$ for each instances $x$, it suffices to apply \eqref{eq:CondProb} to only $ s \in \argmax_{s} P_{Y|S}(x|s) \cup \argmin_{s} P_{Y|S}(x|s)$ for each $x$.
This will also result in a reduction $(1-\alpha)$ of LDP and log-lift, but the design of the randomization scheme and the improvement in data utility need to be studied.
In \cite{Razeghi2020Obfus}, the local information geometry technique is used to approximate the data utility. This paper suggests that it can also be applied to the data privacy. In local proximity $\big| \frac{P_{X|S}(x|s)}{P_X(x)} -1 \big| \leq 1,\forall s,x$, the approximation of LDP in \eqref{eq:PrivApprLDP} can be replaced by the linear equality \eqref{eq:L1Norm}. This treatment is similar to \cite{Huang2017UFS} where the approximation is based on 2nd order Taylor expansion. The linear algebra techniques in \cite{Huang2017UFS} are worth investigating in data privacy.

%
%relative interior, the ball determined by $\ell_2$-norm $\| \frac{\PV_{X=x|S}}{P_{X}(x)} -1 \|_2^2 \leq \frac{\epsilon^2}{P_{X}(x)}$ for each $x \in \X$, where $\PV_{X=x|S} = ( P_{X|S}(x|s) \colon s \in \SA)$. 2nd order Taylor approximation \cite{Huang2017UFS}.

\appendices

\section{Interpretation of Control Variate Method}
\label{app:VARReduct}
%\farhad{[This is very short; why not present it in the main text as a remark?]}
The linear reduction method in~\eqref{eq:CondProb} coincides with the control variate method originally proposed for finding an unbiased estimator in \cite{VarRedContrVar2017}.
It generates new random variable $P_{Y|S}(x|s)$ with the same sample space size as $P_{X|S}(x|s)$, but a strictly smaller variance: for each $x$, $\Var [P_{Y|S}(x|s)]  = \E_S [( P_{Y|S}(x|s) - P_X(x))^2] = (1-\alpha)^2 \Var[P_{X|S}(x|s)] < \Var[P_{X|S}(x|s)],  \ \forall \alpha \in (0,1].$
%\begin{multline*}
%		 \Var [P_{Y|S}(x|s)]  = \E_S [( P_{Y|S}(x|s) - P_X(x))^2] =\\
%		 					 (1-\alpha)^2 \Var[P_{X|S}(x|s)] < \Var[P_{X|S}(x|s)],  \ \forall \alpha \in (0,1].
%\end{multline*}
%\begin{equation*}
%	\begin{aligned}
%		 \Var [P_{Y|S}(x|s)] & = \E_S [( P_{Y|S}(x|s) - P_X(x))^2]\\
%		 					& = \E_S [ (1-\alpha)^2 (P_{X|S}(x|s) - P_X(x))^2]\\
%		 					& = (1-\alpha)^2 \Var[P_{X|S}(x|s)] \\
%                            & < \Var[P_{X|S}(x|s)],  \qquad \forall \alpha \in (0,1].
%	\end{aligned}
%\end{equation*}

\section{Total Variance distance as Utility Loss}
\label{app:DTV}
\begin{figure}[tpb]
	\centering
    \scalebox{0.6}{% This file was created by matlab2tikz v0.4.3.
% Copyright (c) 2008--2013, Nico Schlömer <nico.schloemer@gmail.com>
% All rights reserved.
%
% The latest updates can be retrieved from
%   http://www.mathworks.com/matlabcentral/fileexchange/22022-matlab2tikz
% where you can also make suggestions and rate matlab2tikz.
%
%
% defining custom colors
\definecolor{mycolor1}{rgb}{1,0,1}%
\begin{tikzpicture}

\begin{axis}[%
width=4.5in,
height=2.3in,
scale only axis,
separate axis lines,
xmin=0,
xmax=1,
xlabel={\Large $\alpha$},
ymin=0,
ymax=2.6,
ylabel={\Large utility loss},
legend style={at={(0.02,0.98)},anchor=north west,draw=black,fill=white,legend cell align=left}
]
\addplot [
line width = 1.5pt,
color=blue,
solid,
mark=asterisk,
mark options={solid},
]
table[row sep=crcr]{
0.011 0.0772885887404153\\
0.061 0.320568371675311\\
0.111 0.513210820840208\\
0.161 0.679946671040701\\
0.211 0.828429834667344\\
0.261 0.96248092757314\\
0.311 1.08437528276451\\
0.361 1.19560136290406\\
0.411 1.2971874383319\\
0.461 1.38986528193327\\
0.511 1.47416050097212\\
0.561 1.55044545971092\\
0.611 1.61897108281963\\
0.661 1.67988549226527\\
0.711 1.73324340940055\\
0.761 1.7790079808696\\
0.811 1.81704501651316\\
0.861 1.84710780414823\\
0.911 1.86880770506101\\
0.961 1.8815593447362\\
};
\addlegendentry{$H(X) - I(X;Y)$ for Markov};

\addplot [
line width = 1.5pt,
color=red,
dotted,
mark=o,
mark options={solid},
]
table[row sep=crcr]{
0.011 0.015598\\
0.061 0.086498\\
0.111 0.157398\\
0.161 0.228298\\
0.211 0.299198\\
0.261 0.370098\\
0.311 0.440998\\
0.361 0.511898\\
0.411 0.582798\\
0.461 0.653698\\
0.511 0.724598\\
0.561 0.795498\\
0.611 0.866398\\
0.661 0.937298\\
0.711 1.008198\\
0.761 1.079098\\
0.811 1.149998\\
0.861 1.220898\\
0.911 1.291798\\
0.961 1.362698\\
};
\addlegendentry{$\DTV(X;Y)$  for Markov};

\addplot [
line width = 1.5pt,
color=green,
solid,
mark=star,
mark options={solid},
]
table[row sep=crcr]{
0.011 0.0238153509666119\\
0.061 0.09534475755104\\
0.111 0.149611611860761\\
0.161 0.195985654104696\\
0.211 0.262929517869029\\
0.261 0.287100134448273\\
0.311 0.319648609967296\\
0.361 0.371365189724089\\
0.411 0.43247761380828\\
0.461 0.431359438931822\\
0.511 0.4196760160283\\
0.561 0.47066636465282\\
0.611 0.524712253064937\\
0.661 0.551025633207155\\
0.711 0.529289224244905\\
0.761 0.549141814584803\\
0.811 0.55505644064243\\
0.861 0.571696246155406\\
0.911 0.675360033508267\\
0.961 0.648378620233483\\
};
\addlegendentry{$H(X) - I(X;Y)$ for Non-Markov};

\addplot [
line width = 1.5pt,
color=mycolor1,
dotted,
mark=diamond,
mark options={solid},
]
table[row sep=crcr]{
0.011 0.00461999999999996\\
0.061 0.02562\\
0.111 0.04662\\
0.161 0.06762\\
0.211 0.08862\\
0.261 0.10962\\
0.311 0.13062\\
0.361 0.15162\\
0.411 0.17262\\
0.461 0.19362\\
0.511 0.21462\\
0.561 0.23562\\
0.611 0.25662\\
0.661 0.27762\\
0.711 0.29862\\
0.761 0.31962\\
0.811 0.34062\\
0.861 0.36162\\
0.911 0.38262\\
0.961 0.40362\\
};
\addlegendentry{$\DTV(X;Y)$ for Non-Markov};

\end{axis}
\end{tikzpicture}% 
}
	\caption{The utility losses $H(X) - I(X;Y)$ and $\DTV(X;Y)$ are increasing with $\alpha$. The mutual information $I(X;Y)$ and total variance distance $\DTV(X;Y)$ are determined by the Markov solution~\eqref{eq:MarkovSolution} and non-Markov solution~\eqref{eq:SolutionNonMarkov} for the dataset in Example~\ref{ex:main}.}
	\label{fig:DTVvsMI}
\end{figure}
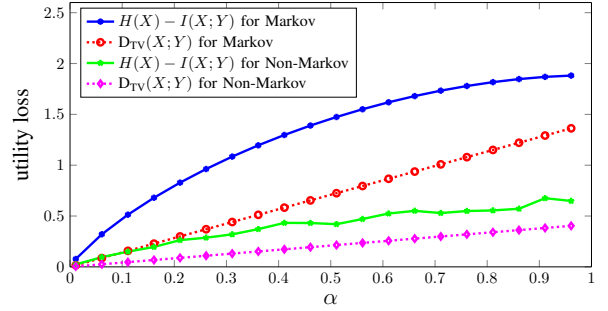

For $Y$ such that $|\Y| = |\X|$, the following transition probability maximizes the mutual information $I(X;Y)$
\begin{equation}
P_{Y|X}^*(x | x') = \begin{cases}
        1 &  x = x' \\
        0 &  x \neq x'
   \end{cases}
\end{equation}
Consider the total variance distance\footnote{Total variance distance is the $f$-divergence $D_f(p\|q) = \sum_{x} q(x) f(\frac{p(x)}{q(x)})$ for $f(t) = \frac{1}{2} |t-1|$. The total variance distance $\DTV(X,Y)$ is between any $P_{X,Y}(x',x) = P_{Y|X}(x | x') P_{X}(x')$ and the optimizer $P_{X,Y}^*(x',x) = P_{Y|X}^*(x | x') P_{X}(x')$.}
$\DTV(X,Y) = \sum_{x} \sum_{x'} P_X(x') \big| P_{Y|X}(x|x) - P_{Y|X}^*(x | x) \big| = 1-\sum_{x} P_{X}(x)P_{Y|X}(x|x)$.
%%
%\begin{equation} \nonumber
%\begin{aligned}
%    \DTV(X,Y) & = \sum_{x} \sum_{x'} p_X(x') \big| p_{Y|X}(x|x) - p_{Y|X}^*(x | x) \big|\\
%%    & = \sum_{x} \Big( p_X(x) \left( 1- p_{Y|X}(x|x) \right) \\
%%    & \qquad + p_{Y}(x) - p_{X}(x) p_{Y|X}(x|x) \Big) \\
%    & =\sum_{x} p_{X}(x) (1-p_{Y|X}(x|x)).
%\end{aligned}
%\end{equation}
%%
It can be seen from Fig.~\ref{fig:DTVvsMI} that $\DTV(X,Y)$ is almost order reversing, i.e., if $I(X;Y) \geq I(X;Y')$, then $\DTV(X;Y) \leq \DTV(X;Y)$. Therefore, for $I(X;Y)$ being a utility measure, $\DTV(X;Y)$ denotes the utility loss.

\section{Proof of Propositions~\ref{prop:DTV} and \ref{prop:ED}}
\label{app:prop:DTV_ED}

    For $U(X;Y) = -\DTV(X;Y)$, the problem \eqref{eq:PUT} is equivalent to $\max \sum_{s,x} P_{Y|S,X}(x|s,x) P_{S,X}(s,x)$ subject to \eqref{eq:PUTconstr}.
    This LP is \emph{separable} in $s$ \cite{Boyd2004Convex}, the maximizer of which can be determined by solving
    \begin{equation} \label{eq:auxProb1}
        \max \sum_{x} P_{Y|S,X}(x|s,x) P_{S,X}(s,x), \text{ s.t. } \eqref{eq:PUTconstr}, \forall x
    \end{equation}
    for each $s$.
    We show below this problem is also separable in $x$.
    For each $s$, rewrite \eqref{eq:PUTconstr} as
    \begin{multline} \label{eq:aux1}
        P_{Y|S,X}(x|s,x) = 1-\alpha \Big( 1-\frac{P_{X}(x)}{P_{X|S}(x|s)} \Big) - \\
        \sum_{x' \colon x' \neq x} P_{Y|S,X}(x|s,x') \frac{P_{X|S}(x'|s)}{P_{X|S}(x|s)}, \quad \forall x.
    \end{multline}
    Since $P_{Y|S,X}(x|s,x) \leq 1$, we rewrite \eqref{eq:aux1} as the inequality
    $ \sum_{x' \colon x' \neq x} P_{Y|S,X}(x|s,x') \frac{P_{X|S}(x'|s)}{P_{X|S}(x|s)} \geq -\alpha \big( 1-\frac{P_{X}(x)}{P_{X|S}(x|s)} \big), \forall x; $
    On the other hand, because $P_{Y|S,X}(x|s,x') \geq 0, \forall x' \neq x$, we have
    $\sum_{x' \colon x' \neq x} p_{Y|S,X}(x|s,x') \frac{P_{X|S}(x'|s)}{P_{X|S}(x|s)} \geq \max \left\{ -\alpha \big( 1-\frac{P_{X}(x)}{P_{X|S}(x|s)} \big),0 \right\},\forall x.$
%    \begin{multline*}
%        \sum_{x' \colon x' \neq x} p_{Y|S,X}(x|s,x') \frac{P_{X|S}(x'|s)}{P_{X|S}(x|s)} \\
%        \geq \max \left\{ -\alpha \big( 1-\frac{P_{X}(x)}{P_{X|S}(x|s)} \big),0 \right\}, \quad \forall x.
%    \end{multline*}
     Apply this inequality to \eqref{eq:aux1} to convert the constraint $\eqref{eq:PUTconstr}, \forall s$ in \eqref{eq:auxProb1} to
     \begin{equation} \label{eq:aux2}
         P_{Y|S,X}(x|s,x) \leq \min \Big\{  1 - \alpha \big( 1 - \frac{P_{X}(x)}{P_{X|S}(x|s)} \big), 1 \Big\}, \forall x.
    \end{equation}
%    \begin{equation} \label{eq:aux2}
%    \begin{aligned}
%        & P_{Y|S,X}(x|s,x) \\
%        &\leq  1-\alpha \Big( 1-\frac{P_{X}(x)}{P_{X|S}(x|s)} \Big) +
%        \min \Big\{ \alpha \big( 1-\frac{P_{X}(x)}{P_{X|S}(x|s)} \big),0 \Big\}  \\
%        &= \min \Big\{  1 - \alpha \big( 1 - \frac{P_{X}(x)}{P_{X|S}(x|s)} \big), 1 \Big\}
%    \end{aligned}
%    \end{equation}
    Then, problem~\eqref{eq:auxProb1} is decomposable in $x$. For each $s$ and $x$, the solution to $\max P_{Y|S,X}(x|s,x) P_{S,X}(s,x), \text{ s.t. } \eqref{eq:aux2}$ is $P_{Y|S,X}^*(x|x,s) = \min \big\{  1 - \alpha \big( 1 - \frac{P_{X}(x)}{P_{X|S}(x|s)} \big), 1 \big\}$, where, by constraint~\eqref{eq:PUTconstr} and $\sum_{x} P_{Y|S,X}(x|s,x') = 1$, we have \eqref{eq:SolutionNonMarkov2} and \eqref{eq:SolutionNonMarkov3}, respectively.
    From \eqref{eq:PUTconstr}, \eqref{eq:SolutionNonMarkov1} and \eqref{eq:SolutionNonMarkov2}, we have \eqref{eq:SolutionNonMarkov4}; For \eqref{eq:SolutionNonMarkov3} and $\sum_{x \in \X} P_{Y|S,X}^*(x|s,x') = 1$, we have \eqref{eq:SolutionNonMarkov5}.

For $U(X;Y) = -\E [d(X,Y)]$, problem \eqref{eq:PUT} is also separable in $s$. From~\eqref{eq:aux2}, we have the constraint $ \sum_{x' \colon x' \neq x} P_{Y|S,X}(x|s,x') \leq \max \big\{ 0, - \alpha \big( 1 - \frac{P_{X}(x)}{P_{X|S}(x|s)} \big) \big\}$, where the objective function $\sum_{x,x' \colon x' \neq x} P_{Y|S,X}(x|s,x') P_{S,X}(s,x')) d(x',x)$ is minimized when $\sum_{x' \colon x' \neq x} P_{Y|S,X}(x|s,x') = \max \big\{ 0, - \alpha \big( 1 - \frac{P_{X}(x)}{P_{X}(x|s)} \big) \big\}$ by the optimizer in \eqref{eq:SolutionNonMarkov}, where the value of $P_{Y|S,X}(x|s,x')$ for all $x \in \X^+(s)$ and $x' \in \X^-(s)$ is determined by \eqref{eq:SolutionNonMarkovED}. \hfill\IEEEQED

\bibliographystyle{IEEEtran}
\bibliography{LDP}

%\newpage\phantom{blabla}
%\newpage

\section{Approximation in~\eqref{eq:PrivAppr}} \label{app:eq:PrivAppr}
Using $\log(1+x) \approx x$, we have the local differential privacy
\begin{equation}\nonumber
    \begin{aligned}
        \LDP (S \rightarrow X)
        &= \max_{x,s,s'} \log \frac{P_{X|S}(x|s)}{P_{X|S}(x|s')} \\
        &= \max_{x,s,s'} \left\{ \log \frac{P_{X|S}(x|s)}{ P_X(x)}  - \log \frac{P_{X|S}(x|s')}{P_X(x)} \right\} \\
        &\approx \max_{x,s,s'} \frac{P_{X|S}(x|s) - P_{X|S}(x|s')}{P_{X}(x)}.
    \end{aligned}
\end{equation}
Similarly,
\begin{equation} \nonumber
    \begin{aligned}
        \LDP (S \rightarrow Y)
        &\approx \max_{x,s,s'} \frac{P_{Y|S}(x|s) - P_{Y|S}(x|s')}{P_{Y}(x)} \\
        &= \max_{x,s,s'} \frac{(1-\alpha) \big( P_{X|S}(x|s) - P_{X|S}(x|s') \big)}{P_{X}(x)} \\
        &= (1-\alpha) \max_{x,s,s'} \frac{P_{X|S}(x|s) - P_{X|S}(x|s')}{P_{X}(x)} \\
        &\approx \LDP (S \rightarrow X).
    \end{aligned}
\end{equation}
For the log-lift, we have
\begin{equation}\nonumber
    \begin{aligned}
        \LogLift (S \rightarrow X) &= \max_{x,s} \left| \log \frac{P_{X|S}(x|s)}{P_X(x)} \right| \\
        &\approx \max_{x,s} \left| \frac{P_{X|S}(x|s)}{P_X(x)} - 1 \right|
    \end{aligned}
\end{equation}
so that
\begin{equation}\nonumber
    \begin{aligned}
        \LogLift (S \rightarrow Y) &= \max_{x,s} \left| \log \frac{(1-\alpha)P_{X|S}(x|s) + \alpha P_X(x)}{P_Y(x)} \right| \\
        &= \max_{x,s} \left| \log \frac{(1-\alpha)P_{X|S}(x|s)}{P_X(x)} + \alpha \right|  \\
        &\approx \max_{x,s} \left| \frac{(1-\alpha)P_{X|S}(x|s)}{P_X(x)} + \alpha - 1 \right|  \\
        &= (1-\alpha) \max_{x,s} \left| \frac{P_{X|S}(x|s)}{P_X(x)} - 1 \right| \\
        &\approx (1-\alpha) \LogLift (S \rightarrow X).
    \end{aligned}
\end{equation}

\section{Proof of Lemma~\ref{lem:Markov}} \label{app:lem:Markov}
    For Markov transition probability $P_{Y|X}(x|x') = P_{Y|S,X}(x|s,x')$ for all $s$, we have $P_{Y|S}(x|s) = \sum_{x} P_{Y|X}(x|x') P_{X}(x'|s)$ for all $s$ and $x$.
    Let $\PMat_{Y|X}$ denote a $|\X| \times |\Y|$ transition probability matrix:
    $ \PMat_{Y|X} = [\PV_{Y=x_1|X} \ \PV_{Y=x_2|X}\ \dotsc  ]$
    where $\PV_{Y=x|X}$ is a column vector such that
    $$ \PV_{Y=x|X} = [P_{Y|X}(x|x') \colon x' \in \X]^\intercal. $$
    Let $\PMat_{X|S}$ be a $|\SA| \times |\X|$ matrix such that $\PMat_{X|S} = [\PV_{X=x_1|S} \ \PV_{X=x_2|X}\ \dotsc ]$, where $ \PV_{X=x|S} = [P_{X|S}(x|s) \colon x \in \X]^\intercal$.
    We define $\PMat_{Y|S}$ and $\PV_{Y=x|S}$ in the same way.
    Rewrite \eqref{eq:CondProb} in column vector form:
    \begin{equation} \label{eq:MarkovSolution}
        \PV_{Y = x|S} = \PMat_{X|S} \PV_{Y = x|X} = (1-\alpha) \PV_{X = x|S} + \alpha P_{X}(x) \One ,
    \end{equation}
    where $\One = [1,\dotsc,1]^\intercal$ denotes an all one column vector.
    For $|\SA| \geq |\X|$, let $\AMat$ be the left inverse matrix of $\PMat_{X|S}$.\footnote{The underlying assumption here is that $\PMat_{X|S}$ is full rank. However, \eqref{eq:SolutionMarkovProof} ensures that the transition probability in Lemma~\ref{lem:Markov} is the Markov solution to any $\PMat_{X|S}$. }
    Then,
    \begin{equation} \label{eq:CondProbMarix}
        \PV_{Y = x|X} = \AMat \PV_{Y = x|S} = (1-\alpha) \AMat \PV_{X = x|S} + \alpha P_{X}(x) \AMat \One.
    \end{equation}
    Let $a_{x,s}$ be the $x$th row and $s$th column entry of $\AMat$ and $\AV_x = [a_{x,s_1} \ a_{x,s_2} \ \dotsc]$ be the row vector of $\AMat$. % such that $\AMat = [\AV_{x_1}^{\intercal} \ \AV_{x_2}^{\intercal} \ \dotsc]^\intercal$.
    Denote $\Eye_{m}$ the identity matrix of dimension $m$.
    From $\AMat \PMat_{X|S} = \Eye_{|\X|}$, we have
    \[ \AV_{x} \PV_{X = x'|S} = \sum_{s}  a_{x,s} P_{X|S}(x'|s) =  \begin{cases}
                                1 & x' = x\\
                                0 & x' \neq x
                             \end{cases}.
    \]
    From $\AMat \PMat_{X|S} \One = \One$, we have
    $ \AV_x \PMat_{X|S} \One = \sum_{s \in \SA} a_{x,s}  \big( \sum_{x' \in \X} p(x'|s) \big)
        = A_x \One = 1, \forall x.  $
    That is $\AMat \One = \One $. We rewrite \eqref{eq:CondProbMarix} as
    $ \PV_{Y = x|X}= (1-\alpha) \Unit_x + \alpha P_X(x) \One, \forall x, $
    where $\Unit_x$ is a unit vector such that the $x$-dim is 1 and all others are zero.
    It is shown in the proof of Lemma~\ref{lem:Markov} that this is in fact the Markov solution for any $\PMat_{X|S}$.  \hfill \IEEEQED

\section{Rank Deficiency of \eqref{eq:SolutionNonMarkov4} and \eqref{eq:SolutionNonMarkov5}}
\label{app:NullSpace}
From \eqref{eq:SolutionNonMarkov5}, we have
\begin{multline*}
    P^*_{Y|S,X}(x|s,x') = \\ - \sum_{\tilde{x} \in \X^+(s) \colon \tilde{x} \neq x} P^*_{Y|S,X}(\tilde{x}|s,x') + \alpha \big( 1- \frac{P_X(x')}{P_{X|S}(x'|s)} \big)
\end{multline*}
for each $x \in \X^+(s)$ and $x' \in \X^-(s)$. Substituting to \eqref{eq:SolutionNonMarkov4}, we get
\begin{multline*}
    \sum_{x' \in \X^-(s)} \Bigg( -\sum_{\tilde{x} \in \X^+(s) \colon \tilde{x} \neq x} P^*_{Y|S,X}(\tilde{x} | s,x') P_{X|S}(x'|s) \\
    + \alpha \Big( P_{X|S}(x'|s) - P_{X}(x') \Big) \Bigg)  = -\alpha \Big( P_{X|S}(x|s) - P_{X}(x) \Big)
\end{multline*}
for each $x \in \X^+(s)$. Reorganize this equality as
\begin{equation}
     \begin{aligned}
        &\sum_{x' \in \X^-(s)} \sum_{\tilde{x} \in \X^+(s) \colon \tilde{x} \neq x} P^*_{Y|S,X}(\tilde{x} | s,x') P_{X|S}(x'|s)\\
        &= \alpha \Bigg[ \sum_{x' \in \X^-(s)} \Big( P_{X|S}(x'|s) - P_{X}(x') \Big) \\
        & \qquad + \Big( P_{X|S}(x|s) - P_{X}(x) \Big)  \Bigg] \\
        &= \alpha \Bigg[ \sum_{x' \in \X^-(s)} \Big( P_{X|S}(x'|s) - P_{X}(x') \Big) \\
        & \qquad + \sum_{x'' \in \X^+(s)} \Big( P_{X|S}(x''|s) - P_{X}(x'') \Big)  \\
        & \qquad - \sum_{\tilde{x} \in \X^+(s) \colon \tilde{x} \neq x} \Big( P_{X|S}(\tilde{x}|s) - P_{X}(\tilde{x}) \Big) \Bigg]\\
        &= -\alpha \sum_{\tilde{x} \in \X^+(s) \colon \tilde{x} \neq x} \Big( P_{X|S}(\tilde{x}|s) - P_{X}(\tilde{x}) \Big).
     \end{aligned}
\end{equation}
This is exactly the resulting equality by summing both sides of \eqref{eq:SolutionNonMarkov4} over all $\tilde{x} \in \X^+(s)$ such that $\tilde{x} \neq x$. Therefore, the dimension of the null space of \eqref{eq:SolutionNonMarkov4} and \eqref{eq:SolutionNonMarkov5} is no less than 1, i.e., \eqref{eq:SolutionNonMarkov4} and \eqref{eq:SolutionNonMarkov5} do not form linear equations with full rank and therefore the resulting constraint set is not singleton.

%%%%%%
%% To balance the columns at the last page of the paper use this
%% command:
%%
%\enlargethispage{-1.2cm}
%%
%% If the balancing should occur in the middle of the references, use
%% the following trigger:
%%
\IEEEtriggeratref{3}
%%
%% which triggers a \newpage (i.e., new column) just before the given
%% reference number. Note that you need to adapt this if you modify
%% the paper.  The "triggered" command can be changed if desired:
%%
%\IEEEtriggercmd{\enlargethispage{-20cm}}
%%
%%%%%%

%%%%%%
%% References:
%% We recommend the usage of BibTeX:
%%
%\bibliographystyle{IEEEtran}
%\bibliography{definitions,bibliofile}
%%
%% where we here have assume the existence of the files
%% definitions.bib and bibliofile.bib.
%% BibTeX documentation can be obtained at:
%% http://www.ctan.org/tex-archive/biblio/bibtex/contrib/doc/
%%%%%%

%% Or you use manual references (pay attention to consistency and the
%% formatting style!):

\end{document}